\documentclass{ws-p8-50x6-00}

\begin{document}

\title{Experimental evidence for spinodal decomposition in
multifragmentation of heavy systems}

\author{G.~T\u{a}b\u{a}caru$^{1,2}$, B.~Borderie$^1$, 
\underline{M.F.~Rivet}$^1$,
P.~Chomaz$^3$,M.~Colonna$^4$, J.D.~Frankland$^3$, A.~Guarnera$^4$,
M.~P\^arlog$^2$, B.~Bouriquet$^3$,
A.~Chbihi$^3$, S.~Salou$^3$, J.P.~Wieleczko$^3$,
G.~Auger$^3$, Ch.O.~Bacri$^1$, N.~Bellaize$^5$, R.~Bougault$^5$,
 R.~Brou$^5$, P.~Buchet$^6$, J.L.~Charvet$^6$,
J.~Colin$^5$,D.~Cussol$^5$, R.~Dayras$^6$, A.~Demeyer$^7$, D.~Dor\'e$^6$,
D.~Durand$^5$, E.~Galichet$^{1,8}$, E.~Gerlic$^7$, B.~Guiot$^3$, S.~Hudan$^3$,
D.~Guinet$^7$, P.~Lautesse$^7$, F.~Lavaud$^1$, J.L.~Laville$^3$,
J.F.~Lecolley$^5$, C.~Leduc$^7$, R.~Legrain$^6$, N.~Le Neindre$^5$,
O.~Lopez$^5$, M.~Louvel$^5$, J.~{\L}ukasik$^1$, A.M.~Maskay$^7$,
L.~Nalpas$^6$,
J.~Normand$^5$, P.~Paw{\L}owski$^1$, E.~Plagnol$^1$, E.~Rosato$^9$, 
F.~Saint-Laurent$^3$\footnote{\lowercase{present address:} DRFC/STEP, 
CEA/C\lowercase{adarache}, F-13018
S\lowercase{aint}-P\lowercase{aul}-\lowercase{lez}-D\lowercase{urance
Cedex}, F\lowercase{rance}.},
J.C.~Steckmeyer$^5$, B.~Tamain$^5$, L.~Tassan-Got$^1$,
E.~Vient$^5$, C.~Volant$^6$\\
 (INDRA collaboration)
}

\address{$1$ Institut de Physique Nucl\'eaire, IN2P3-CNRS, F-91406 Orsay Cedex,
 France.~\\
$2$ National Institute for Physics and Nuclear Engineering, RO-76900
Bucharest-M\u{a}gurele, Romania.~\\
$3$ GANIL, CEA et IN2P3-CNRS, B.P.~5027, F-14076 Caen Cedex, France.~\\
$4$ Laboratorio Nazionale del Sud, Viale Andrea Doria, I-95129 Catania,
Italy.~\\
$5$ LPC, IN2P3-CNRS, ISMRA et Universit\'e, F-14050 Caen Cedex, France.~\\
$6$ DAPNIA/SPhN, CEA/Saclay, F-91191 Gif sur Yvette Cedex, France.~\\
$7$ Institut de Physique Nucl\'eaire, IN2P3-CNRS et Universit\'e,
F-69622 Villeurbanne Cedex, France. \\
$8$ Conservatoire National des Arts et M\'etiers, F-75141 Paris Cedex 03.~\\
$9$ Dipartimento di Scienze Fisiche e Sezione INFN, Universit\a di Napoli
`Federico II'', I-80126 Napoli, Italy.}


\maketitle

\abstracts{
Multifragmentation of fused systems was observed for central collisions
between 32 AMeV $^{129}$Xe and Sn, and 36 AMeV $^{155}$Gd and U.
Previous extensive comparisons between the two systems led to the hypothesis 
of spinodal decomposition of finite systems as the origin of 
multifragmentation for incident energies around 30 AMeV. New
results on velocity and charge correlations of fragments bring strong
arguments in favor of this interpretation.}

\section{Introduction}

The decay of highly excited nuclear systems through multifragmentation is
currently a subject of great interest in nucleus-nucleus collisions. The
recent advent of powerful 4$\pi$ devices brought high quality experimental
data which allow a careful selection of well defined fused systems
undergoing multifragmentation. We present in this contribution new results
on the multifragmentation of heavy fused systems, formed in central
32 AMeV $^{129}$Xe+$^{nat}$Sn, and 36 AMeV $^{155}$Gd+U collisions. 
Details on the experiment, performed at GANIL with the INDRA
4$\pi$ array, and on event selection can be found in 
ref~\cite{indra,FR98,RI98}.

\begin{figure}[htb]
\noindent \includegraphics[width=6cm]{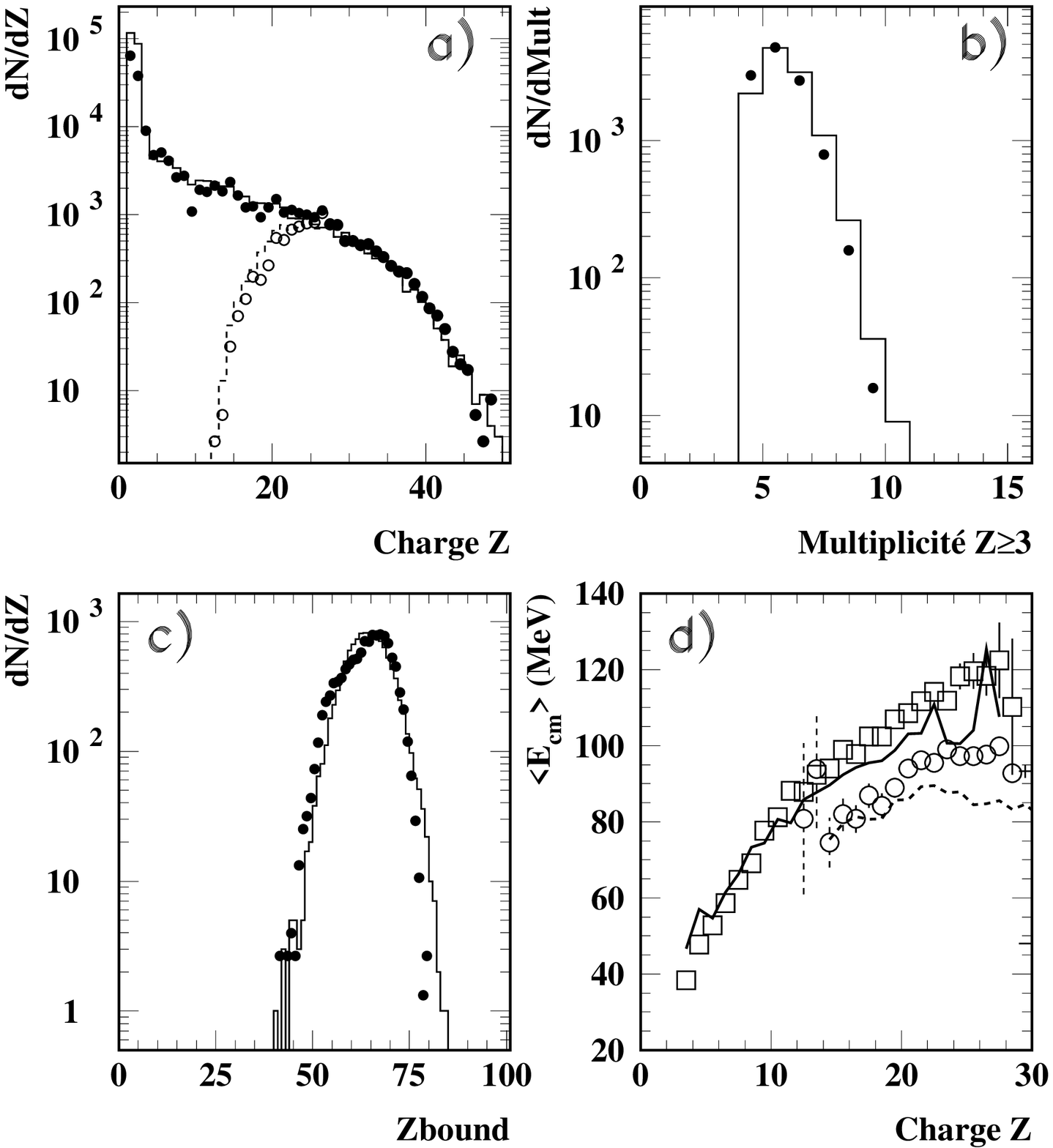}
\vspace*{-6cm} \begin{flushright} 
\includegraphics[width=6cm]{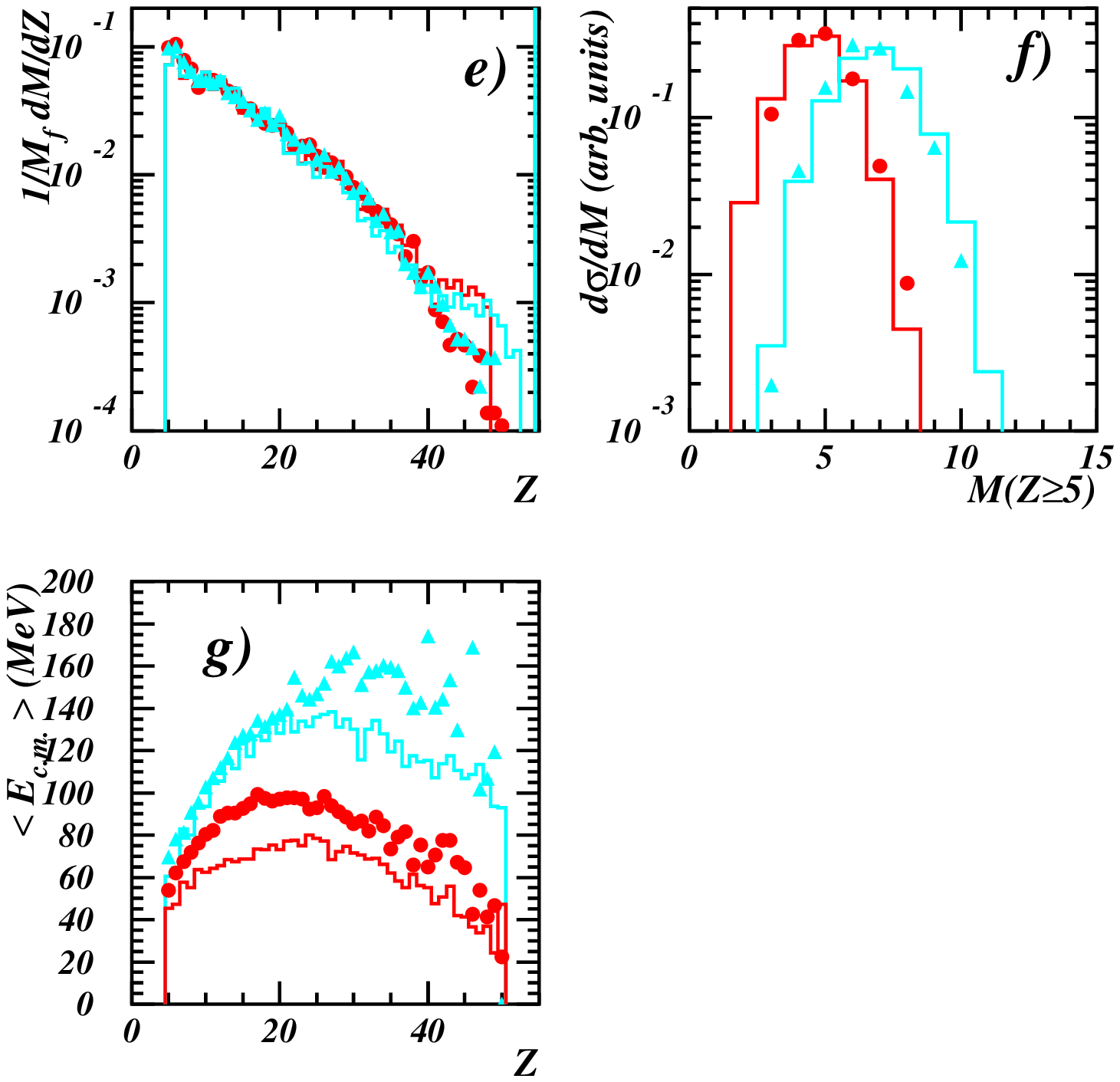}
\end{flushright}
\caption{Comparison of experimental data with SMM  (a-d) and BOB (e-g)
simulations.  a-d concerns the 32 AMeV Xe+Sn system, the lines are for data 
and the symbols for SMM (all fragments, except open circles and dashed lines 
which refer to the largest fragment
of each partition). Z$_{bound}$ represents the sum of the charges of all 
fragments. In e-g the symbols are for data and the lines for BOB
simulations. Light grey lines and triangles stand for Gd+U and black lines
and circles for Xe+Sn.}
\label{comp}
\end{figure}

First evidence for a bulk effect was found in the comparison of the two
multifragmenting systems, which have the same available energy $\sim$ 7 AMeV:
the charge distributions are identical while the number of fragments scales
as the charge of the systems\cite{RI98,FR98}. This scaling could simply be
the sign of the breaking of statistically equilibrated systems, but could
also be envisaged as the occurrence of spinodal decomposition
of very heavy composite systems (A$>$250).
Indeed if systems break in the spinodal region of the (T,$\rho$) plane ,  
a ``primitive'' break-up into fragments with 
a favoured size is predicted~\cite{GU96}, in connexion with the wave 
lengths of the most unstable modes in nuclear matter.
The enhancement of equal-sized fragment partitions
is however washed out by several effects (beating of different modes, 
coalescence of nuclear-interacting fragments, finite size of the systems, 
and secondary decays).  Stochastic mean-field simulations (BOB)~\cite{GU97}
which describe the entire collision process up to the final fragment
de-excitation, well account for fragment multiplicity and charge distributions
as well as for the  average fragment kinetic energies (fig.~\ref{comp}e-g).
The SMM statistical model~\cite{BO95}  however also well reproduces the same 
distributions,
provided that the mass, charge and excitation energy of the system are traced
back to the ``freeze-out'' time, where fragments cease to feel the nuclear
interaction~\cite{NLN99,BOU00} (fig.~\ref{comp}a-d). Therefore the variables 
considered above do
not furnish a stringent enough test of a spinodal decomposition, and 
one has to turn to correlation studies.
In the following sections the samples of experimental and simulated (BOB, SMM)
events are  exactly the same as those from which fig~\ref{comp}a-g were built. 

\section{Velocity correlations}

\begin{figure}[htb]
\noindent \includegraphics[width=7cm]{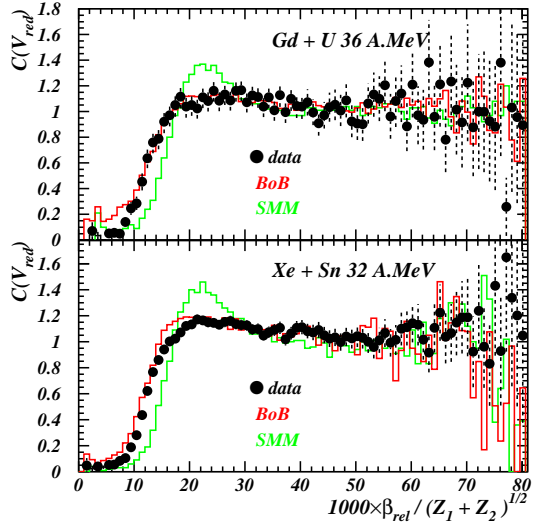}
\vspace*{-4cm} \begin{flushright} \begin{minipage}{4.5cm}
\caption{Correlation functions for the reduced fragment  relative
velocities. The fragment charges are in the range 5-20. Experimental data
(points) are compared with the results of dynamical simulations and of the
statistical model SMM, for the Gd+U (top) and Xe+Sn (bottom)
multifragmenting events.} \label{correlV}
\end{minipage} \end{flushright}
\end{figure}

Reduced fragment-fragment velocity correlations were built,
 mixing pairs of fragments with different charges to increase statistics. 
The same procedure was used for experimental data and for simulated events.
These correlation functions should reflect the topology of the events at
freeze-out, and especially the strength of the Coulomb repulsion.
Experimental and calculated (BOB and SMM)
velocity correlation functions are shown in fig~\ref{correlV} for the
two systems under study. In this figure the fragment are limited to the 
range Z=5-20, but similar pictures are obtained when the
heavier fragments are included (for both systems $<Z_{max}> \sim25$).
In both cases the BOB simulations perfectly match the
experimental data, indicating that the configuration at freeze-out is
correctly described. As this topology is not of bubble-type,it rules out 
surface instabilities as being the cause of multifragmentation.
Conversely, the SMM calculations, which in this version simply
randomly place the fragment in a volume equal to 3 times the normal nuclear
volume, do not reproduce the velocity correlations, in particular the
Coulomb depletion is too broad, and a bump appears just after. In
ref~\cite{SCH94} such a bump  marks  the presence of one very big
fragment in the partitions. It is not the case here as it was shown that 
the details of the partitions  are well reproduced by this SMM calculation. 

\section{Charge correlations}

\begin{figure}[htb]
\noindent \includegraphics[width=7cm]
{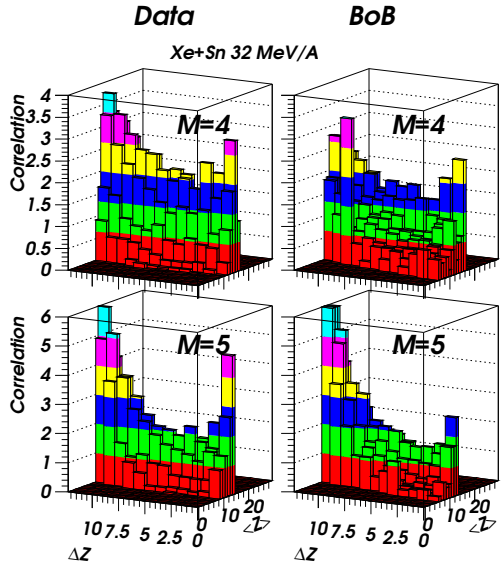}
\vspace*{-3cm} \begin{flushright} \begin{minipage}{4.5cm}
\caption{Charge correlations for events with fragment multiplicities 4 (top)
and 5 (bottom). The left column refers to experimental data, and the right
column to results of the dynamical simulation.}\label{correlZ}
\end{minipage} \end{flushright}
\end{figure}

A compact way of describing event partitions is to build higher order charge
correlation functions, defined as the ratio of yields Y($<$Z$>$,$\Delta$Z)
for correlated and decorrelated fragments; $<$Z$>$ is the average fragment
charge and $\Delta$Z the standard deviation {\em per event}~\cite{MO96}.
In such a
representation, any privileged partition (down to the level of 0.1\%) will
appear as a peak at the corresponding $<$Z$>$,$\Delta$Z bin.
Charge correlation functions built for experimental  and BOB simulated
events are shown in fig.~\ref{correlZ}; only for the Xe+Sn system  was the 
statistics
high enough for such a study. For all fragment multiplicities (we show here
only M=4 and 5) the experimental charge correlation has a peak in the bin
$\Delta$Z = 0-1, indicating an enhancement of partitions with fragments of
equal size. This peak corresponds to a constant value of the product
M$\times <$Z$>$. Its existence is specific of the fused events selected,
it does not appear for other very dissipative types of collisions~\cite{BOR00}.
The number of events in the peaks at $\Delta$Z = 0-1 summed over all 
multiplicities
corresponds to 0.1\% of the fused events. The same features are observed in
events simulated with BOB (fig.~\ref{correlZ} right). Although in BOB we know
that all events multifragment through spinodal decomposition, only 0.1\%
of them survive the blurring of the favored initial fragment size. Thus we
state that all the experimentally selected fused systems also underwent 
spinodal decomposition. Finally the events simulated with the version of SMM
considered here do not present any partition with equal-sized
fragments~\cite{BOR00}.

\section{Conclusions}

New detailed analyses of fragment velocity and charge correlation functions
for very heavy multifragmenting systems strongly support the previous
assumption that spinodal decomposition is at the origin of
multifragmentation, for very heavy systems at incident energies around 
30-35 AMeV.


\begin{thebibliography}{99}
\bibitem{indra}J.~Pouthas et al., 
\Journal{\NIMA}{357}{418}{1995}.
\bibitem{FR98} J.D.~Frankland et al., (INDRA coll.) submitted to {\em Nucl.
Phys.} A.
\bibitem{RI98} M.F. Rivet et al. (INDRA coll.), \Journal{\PLB}{430}{217}{1998}.
\bibitem{GU96} A. Guarnera et al., \Journal{\PLB}{373}{267}{1996}.
\bibitem{GU97} A. Guarnera et al., \Journal{\PLB}{403}{191}{1997}.
\bibitem{BO95}J. Bondorf et al., \Journal{\PRep}{257}{133}{1995} 
\bibitem{NLN99} N. Le Neindre, th\`ese, Universit\'e de Caen, LPCC T 99 02. 
\bibitem{BOU00}
R. Bougault et al.,  (INDRA coll.) Proc. Bormio (2000), p. 404.
\bibitem{SCH94} O. Schapiro and D.H.E. Gross, \Journal{\NPA}{576}{428}{1994}.
\bibitem{MO96} L.G. Moretto et al., \Journal{\PRL}{77}{2634}{1996}.
\bibitem{BOR00}
G.~T\u{a}b\u{a}caru et al.,  (INDRA coll.) Proc. Bormio (2000), p. 433.

\end{thebibliography}
\end{document}